\documentclass[lettersize,journal]{IEEEtran}

\usepackage{amsmath,amssymb,amsfonts}
\usepackage{mathtools}
\usepackage{graphicx}
\usepackage{textcomp}
\usepackage{xcolor}
\usepackage{caption}
\usepackage{subcaption}
\usepackage{float}

\usepackage{biblatex}
\usepackage{tikz}
\usetikzlibrary{positioning,shapes,arrows,arrows.meta}
\tikzstyle{process} = [rectangle, text centered, draw=black,thick, inner sep=5pt]
\tikzstyle{decision} = [diamond, text centered, draw=black,thick]

\usepackage[ruled,linesnumbered]{algorithm2e}
\SetKwBlock{Loop}{Loop}{end}
\SetKwComment{Comment}{// }{}

\DeclarePairedDelimiter\abs{\lvert}{\rvert}

\begin{document}

\title{Federated Learning and Evolutionary Game Model for Fog Federation Formation}

\author{Zyad Yasser,
        Ahmad Hammoud,
        Azzam Mourad,
        Hadi Otrok,
        Zbigniew Dziong,
        and~Mohsen Guizani

\thanks{Corresponding author: A. Hammoud (ahmad.hammoud.1@ens.etsmtl.ca)}
\thanks{Z. Yasser is with the Division of Science, New York University, Abu Dhabi, UAE}
\thanks{A. Hammoud is with the Department of Electrical Engineering, Ecole de Technologie Superieure (ETS), Montreal, Canada. He is also with the Artificial Intelligence \& Cyber Systems Research Center, Lebanese American University, Beirut, Lebanon. In addition, he is with Mohammad Bin Zayed University of Artificial Intelligence, Abu Dhabi, UAE}
\thanks{A. Mourad is with the KU 6G Research Center, Department of CS, Khalifa University, Abu Dhabi, UAE. He is also with the Artificial Intelligence \& Cyber Systems Research Center, Lebanese American University, Beirut, Lebanon.}
\thanks{H. Otrok is with the Center of Cyber-Physical Systems (C2PS), Department of CS, Khalifa University, Abu Dhabi, UAE}
\thanks{Z. Dziong is with the Department of Electrical Engineering, Ecole de Technologie Superieure (ETS), Montreal, Canada}
\thanks{M. Guizani is with the Mohammad Bin Zayed University of Artificial Intelligence, Abu Dhabi, UAE}}


\markboth{}%
{Yasser \MakeLowercase{\textit{et al.}}: Federated Learning and Evolutionary Game Model for Fog Federation Formation}


\maketitle

\begin{abstract}
    In this paper, we tackle the network delays in the Internet of Things (IoT) for an enhanced QoS through a stable and optimized federated fog computing infrastructure. Network delays contribute to a decline in the Quality-of-Service (QoS) for IoT applications and may even disrupt time-critical functions. Our paper addresses the challenge of establishing fog federations, which are designed to enhance QoS. However, instabilities within these federations can lead to the withdrawal of providers, thereby diminishing federation profitability and expected QoS. Additionally, the techniques used to form federations could potentially pose data leakage risks to end-users whose data is involved in the process. In response, we propose a stable and comprehensive federated fog architecture that considers federated network profiling of the environment to enhance the QoS for IoT applications. This paper introduces a decentralized evolutionary game theoretic algorithm built on top of a Genetic Algorithm mechanism that addresses the fog federation formation issue. Furthermore, we present a decentralized federated learning algorithm that predicts the QoS between fog servers without the need to expose users' location to external entities. Such a predictor module enhances the decision-making process when allocating resources during the federation formation phases without exposing the data privacy of the users/servers. Notably, our approach demonstrates superior stability and improved QoS when compared to other benchmark approaches.
\end{abstract}


\section{Introduction}
\label{section:intro}

\IEEEPARstart{S}{ince} the early history of computing, we have been observing computers getting smaller, more powerful, and more widespread. This was clearly observed in the 1980s with the advent of personal computers \cite{abbate1999getting}. More recently, this revolution of computing has been taking place via the Internet of Things (IoT). IoT applications are countless. Starting from smart thermostats to smart cameras, and driverless vehicles, IoT applications have been essential in our lives. 
IoT devices tend to offload their processing data to more powerful external devices such as cloud servers \cite{al2015internet}. However, in some particular cases, cloud providers may not be able to meet the increasing quality of service (QoS) demands of IoT applications. Mainly because some IoT applications require really low latency such as autonomous vehicles, and smart traffic control. To fix this issue, a new technology was conceived, fog computing \cite{hu2017survey}. 

Fog computing is an extension of cloud computing that brings cloud servers to the edge of the network. Having servers at the network edge next to the IoT application has a twofold benefit. On one hand, it reduces the network's latency as fog computing enables real-time data processing and analytics at the network edge, reducing the need for extensive data transmission to centralized cloud servers. On the other hand, it equips the fog servers with location awareness, an essential feature for many IoT applications such as smart vehicles \cite{FogIot}. However, building fog servers demands high deployment costs. Therefore, the fog federation concept has arisen to solve this problem efficiently. Fog federation refers to fog providers sharing their unallocated resources instead of keeping them idle \cite{hammoud2022demand}. Fog federations increase the QoS of the fog servers: offloading tasks to another server might be beneficial during the shortage of resources. Also, offloading could provide lower transmission delay if we choose the target server to be closer to the clients.

Fog federation formation problem is classified as NP-hard \cite{NPHard}, indicating that there is no known algorithm that can solve the problem in polynomial time. Therefore, researchers typically employ heuristics, meta-heuristics, or approximation algorithms to find near-optimal solutions to the problem \cite{hammoud2022dynamic, sharmin2020toward}. The formation algorithm is required to have low complexity, low accuracy and respect the privacy of the user data, while the formed federations must be stable (i.e., providers would not want to deviate from their federations) and profitable to the providers. Developing efficient and effective algorithms for fog federation formation is an ongoing area of research, with the aim of striking a balance between computational complexity, solution accuracy, privacy preservation, and provider profitability. Most literature techniques tended to concentrate on one or two requirements while disregarding the remaining ones. For example, \cite{GTMainFog} did not have a QoS prediction model to improve the accuracy, while \cite{Hala} had a broker which may violate the privacy of the users.

The objective of this work is to devise a (1) stable, (2) profitable, (3) computationally efficient, (4) data privacy-aware, and (5) fully decentralized approach for the formation of the fog federations, without the need for a third party to handle the management. We first address the problem of privacy of collecting users' request metadata by advising federated learning mechanisms that can predict the QoS of the computing infrastructure without compromising the data privacy of its users. A key aspect of our framework is the client selection process that involves identifying specific entities to participate in devising the infrastructure QoS predictor. We explore the strategies employed to address the heterogeneity of clients (fog providers) and their data distributions \cite{fu2023client}. Moreover, we present our approach to client selection, which leverages a weighted algorithm, ensuring that clients with varying dataset sizes contribute optimally to the global model. Our research sheds light on the critical role of client selection in enhancing the efficiency, accuracy, and privacy preservation of federated learning systems, providing insights into the advancements made in this evolving field. Subsequently, we trained and assessed this model against other benchmark models, and then employed it as a QoS predictor for our approach. 
Moreover, we investigate an efficient Genetic Algorithm (GA) model to assign federations and compose the infrastructure efficiently. Afterward, boosted by the GA, we employ an evolutionary game theoretical approach to help stabilize the formed set of federations and maintain an adequate service for the users. As a result, the derived architecture using this approach exhibits exceptional metrics in terms of QoS and stability when compared to other methods.

\noindent\\
The main contributions of the paper are as follows:
\begin{itemize}
    \item Advising a novel architecture that utilizes federated learning to predict the QoS in order to maximize the accuracy of the federated formation step. Such a process can guarantee that the privacy of the users’ data, which to the best of our knowledge, has not been addressed in similar works.
    \item Introducing an innovative approach to client selection within the federated learning framework, leveraging a weighted algorithm to optimize client contributions based on varying dataset sizes.
    \item Advising a genetic algorithm technique to optimize the profit, and QoS achieved by the fog federations.
    \item Optimizing the profit and the QoS achieved by fog federations while maintaining the stability of the approach using an evolutionary game theoretic technique.
\end{itemize}

The rest of the paper is organized as follows. Section \ref{section:related_work} reviews the current literature on federation formation techniques. Sections \ref{section:scheme} and \ref{section:problem_and_objectives} present the scheme and the formulation of the problem. Section \ref{section:methodology} explains the methodology used in the paper. Section \ref{section:evaluation} discussed the experimental results for the proposed technique.  Section \ref{section:conclusion} concludes the paper.

\section{Related Work}
\label{section:related_work}
In this section, we mainly focus on presenting the work done on federated computing. First, we start with federated cloud computing as it provides a great perspective on fog computing. Then, we shift to federated fog computing.

In \cite{CloudFed}, the paper discussed several cloud federation formation techniques. The main cloud federation formation mechanism (CFFM) discussed in this paper was designed as a game where agents dynamically cooperate to form the targeted federations. That CFFM was demonstrated to be stable, i.e., the agents won't have any incentive to leave their algorithm-recommended federations. One potential concern with the CFFM is data privacy, which may arise from the requirement that agents share information to form targeted federations. Specifically, the agents may need to share data on their resource availability, utilization, and performance to help the CFFM algorithm recommend suitable federations. Sharing such information may raise concerns related to data privacy, as agents may not be comfortable sharing sensitive information with others. Additionally, sharing such data could potentially expose agents to security risks, such as hacking or data breaches, if the data is not adequately secured or if agents do not trust each other. In \cite{GTMainCloud}, the authors built upon Mashayekhy et al work in \cite{CloudFed} and came up with two new techniques for forming cloud federations. The first technique was the genetic algorithm, a meta-heuristic inspired by the process of natural selection. The genetic algorithm outperformed the algorithm mentioned in \cite{CloudFed}. However, it was converging slowly to the solution. Then, the authors proposed a fast converging algorithm that outperformed the aforementioned algorithms. That algorithm was utilizing evolutionary game theoretic techniques that improved the profit up to 10\% compared to Mashayekhy's method. In \cite{FogIot}, the authors defined fog computing and compared it with the traditional cloud computing models. The main characteristics of fog computing are its edge location, location awareness, and low latency. The paper went on to describe one of the prominent technologies that require fog computing such as connected vehicles, smart traffic lights, and smart grids. The authors in \cite{veillon2019f} focused on video streaming services. They devised a solution for reducing their latency metric. Through federating neighboring fog nodes, the authors leveraged data caching and fetching on the edge of the network. Their solution reduced the latency of the data when being fetched on demand. In \cite{GTMainFog}, the authors adopted the evolutionary game theoretic cloud federation formation technique introduced in \cite{GTMainCloud} to fog federations. However, the model lacks a way to predict the QoS dynamically, a requirement for forming the federations. In \cite{anglano2018game}, the authors employed a hedonic game theoretical approach technique to ameliorate the providers' payoff. Their formation technique measures the resources and profit of the fog providers when making clustering decisions in order to provide a better service quality. In \cite{sharmin2020towards}, the authors proposed a resource management framework that establishes a federated fog acting as a consortium to share free resources among the consortium members. The authors also devised a model based on pricing for sharing the computing resources while, at the same time, limiting the offloading mechanism among units to other members. Most of the problems addressed above were tackled in \cite{Hala}. The authors, Shamseddine et al, devised a machine learning algorithm to predict the QoS accurately ahead of federation formation. However, the privacy issue of the broker having access to the data of the clients remained. This shortcoming could be fixed by using a decentralized learning model instead. Additionally, \cite{Hala} used the genetic algorithm as a federation formation technique which is not as efficient as was proven in the literature \cite{CloudFed}. To address this, we are also exploring this by extending Shamseddine et al model to use evolutionary game theoretic models for federation formation while utilizing a federated learning approach to address the networking data privacy shortcomings.

\section{Scheme Overview}
\label{section:scheme}

\begin{figure*}
    \centering
    \includegraphics[width=1\linewidth]{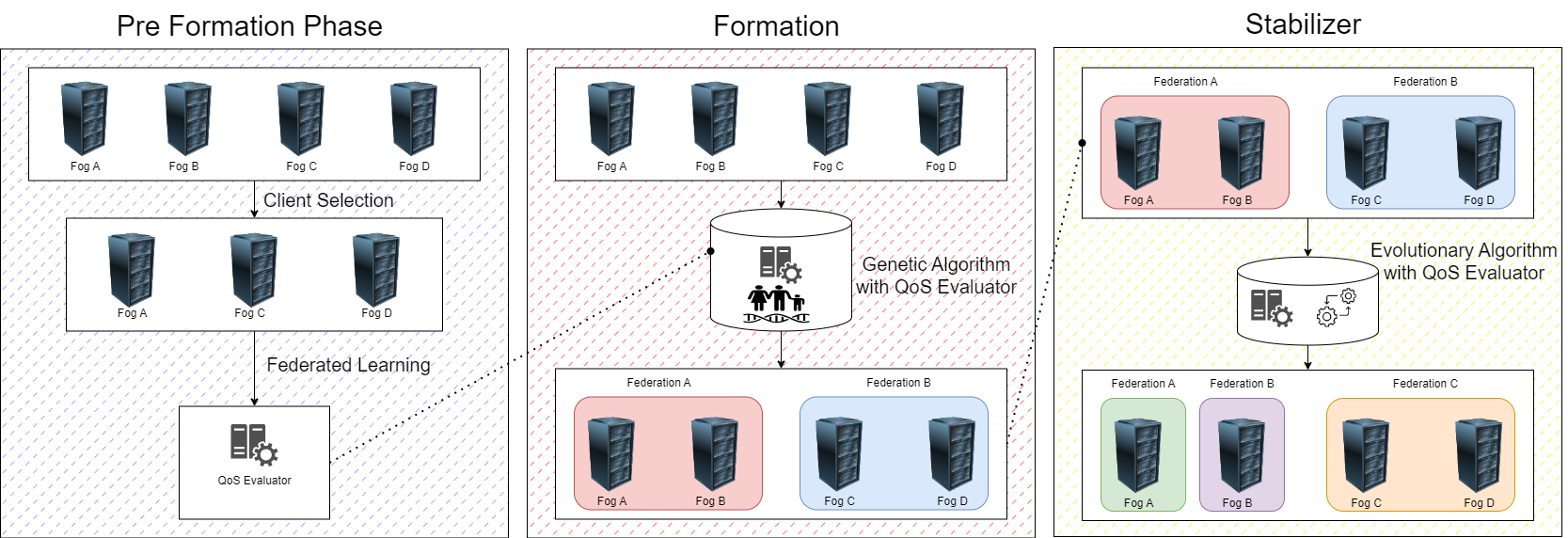}
    \caption{\textmd{Novel Federated Fog Architecture Formation}}
    \label{fig:arch}
\end{figure*}

In this section, we introduce the architecture of the fog federation formation and its mechanism. In addition, we explain the role and duties of each component. 

\subsection{Components}
In IoT context, there are many fog providers each with their own fog servers. Each fog server has its own geographic location and capacity. Due to resource constraints, Application Content Providers are unable to deploy all necessary services on their devices. Consequently, such services are hosted on available fog nodes. Each application content provider is contracted with one or more fog providers to offer resources for its users. Sometimes a fog provider is over-occupied and can’t meet the Application Content Providers’ needs. In this case, it will lead to the dissatisfaction of the user, which can further lead to monetary losses. Federation solves this by delegating the tasks to a fog server of a different provider. We elaborate on each entity in detail.

\textbf{Application Content Providers}: the entities that are willing to offer services to users for economical profit. Application Content Providers cooperate with fog providers to distribute their services on fog servers as needed to provide better QoS for their users.

\textbf{Users}: the primary entities for whom the entire architecture is developed. A user requests a service provided by an application content provider offered by a fog provider on a fog server in a particular location.

\textbf{Fog Providers}: they own fog servers, typically situated in close proximity to end-users to ensure a seamless service experience. Fog providers own several fog nodes, each capable of deploying one or more services offered by an Application Content Provider.

\textbf{Broker}: in charge of optimizing the performance of the proposed architecture. In our proposed model, the broker function will vanish, and a decentralized protocol will substitute it. The broker's mere responsibility would be to collect federated learning models from fog providers/servers and combine them to generate a global model. Then, the broker would send the global model back. The broker will need to operate a privacy-preserving protocol that keeps the dynamic fog federations formation as efficient as possible. The techniques used for fog federation formation and federated learning will be discussed in section \ref{section:methodology}.

\subsection{Mechanism}
We propose a mechanism composed of 3 stages as depicted in Figure \ref{fig:arch}. The first stage is the Pre-Formation Phase. In such phase, we prepare all the necessary data to enable a smooth formation mechanism. Precisely, we initiate a federated learning mechanism coupled with an efficient client selection technique to devise a QoS Evaluator tool. The tool will be able to infer the QoS perceived by any formed federation out of the participating fog providers. The second phase is the formation. In that phase, we adopt a Genetic Algorithm technique that can efficiently form the set of federations while maximizing their potential gain. Lastly, a stabilizer phase takes place. The stabilizer - operated by an evolutionary algorithm theoretical model - will be able to assist the set of federations to reach their stability by relying on the replicator dynamic to achieve the Evolutionary Stable Strategy. In the upcoming sections, we will elaborate further on these three components.

\section{Problem Formulation and Objectives}
\label{section:problem_and_objectives}

\subsection{Problem Formulation}
\label{section:problem}

We consider a setting with a set $P$ of $n$ fog providers,
each $p_i \in P$ has a set of servers $S_{p_i} = \{s_{p_i}^j\}_{j=1}^{n_{p_i}}$. Note that every provider $p_i$ has $n_{p_i}$ servers, $s_{p_i}^j$ where $j$ vary from $1$ to $n_{p_i}$.
With each server $s_{p_i}^j$, there is an associated particular geographical
location $L(s_{p_i}^j)$. 

Let $F=\{f_j\}_{j=1}^{m}$ be the set of $m$ federations to which each $p_i$
consider allocating a subset of its servers. We refer to the set of $n^{f_j}$ servers
allocated to $f_j$ by $S^{f_j}$. Each $p_i$ commits to each federation
$f_j$ by a (possibly empty) set of $n_{p_i}^{f_j}$ servers $S_{p_i}^{f_j} \subseteq S_{p_i}$ such that
\begin{align}
    \bigsqcup_{p_i \in P} S_{p_i}^{f_j} = S^{f_j} \label{eq:1} \\
    \bigsqcup_{f_j \in F} S_{p_i}^{f_j} = S_{p_i} \label{eq:2}
\end{align}
On contrary to the limitations of the other models in the literature ~\cite{GTMainCloud, GTMainFog}, this model treats 
each provider as a divisible good that can be divided into subsets 
and each subset is in a possibly-different federation.

Let $X_{p_i}$ be the set of all possible pure strategies of $p_i$
where $x_i = (S_{p_i}^{f_1}, S_{p_i}^{f_2}, \hdots, S_{p_i}^{f_m}) \in X_{p_i}$
such that $x_i$ respects eq. \eqref{eq:2}. \\
$X = X_1 \times X_2 \times \hdots \times X_n$ be the search space, i.e., the set of all possible states of the system. 

The set of applications allocated to federation $f_i$ is represented by the
set $A_{f_i} = \{a_{f_i}^{1}, a_{f_i}^{2}, \hdots\}$. Likewise, each user $usr_j$
is located at location $L(usr_j)$ and is requesting a set of services/applications
$A_{usr_j} = \{a_{usr_j}^{1}, a_{usr_j}^{2}, \hdots\}$.

We denote the operational cost and the traffic cost of a server $s$ by
$OC(s)$ and $TC(s)$ respectively. We consider the utility function of
a federation $f_i$ to be the revenue minus the costs as follows:
\begin{align}
    u(f_i) &= Rev(f_i) - \sum_{s \in S^{f_i}} \big( OC(s) + TC(s) \big) \\
    Rev(f_i) &= \sum_{a \in A_{f_i}} \sigma_{a,f_i} \cdot payment(a)
\end{align}

\begin{table}[H]
    \centering
    \caption{\textmd{Symbol Descriptions used in Problem Formulation }}
    \normalsize
    \begin{tabular}{p{0.1\linewidth} p{0.75\linewidth}} 
        \hline
        Symbol & Description \\
        \hline
        $P$ & the set of all fog providers  \\[0.075cm]
        $n$ & the number of all fog providers \\[0.075cm]
        $p_i$ & fog provider $i$ \\[0.075cm]
        $S_{p_i}$ & the set of servers of the fog provider $i$ \\[0.075cm]
        $s_{p_i}^j$ & the server $j$ of the fog provider $i$ \\[0.075cm]
        $L(s_{p_i}^j)$ & the location of server $j$ of the fog provider $i$ \\[0.075cm]
        $F$ & the set of all fog federations\\[0.075cm]
        $m$ & the number of all fog federations\\[0.075cm]
        $f_j$ & federation $j$\\[0.075cm]
        $n^{f_j}$ & the number of servers allocated to federation $j$\\[0.075cm]
        $S^{f_j}$ & the set of servers allocated to federation $j$\\[0.075cm]
        $n_{p_i}^{f_j}$ & the number of servers allocated to federation $j$ by provider $i$\\[0.075cm]
        $S_{p_i}^{f_j}$ & the set of servers allocated to federation $j$ by provider $i$ \\[0.075cm]
        $X_{p_i}$ & the set of strategies available to player $i$\\[0.075cm]
        $X$ & the search space\\[0.075cm]
        $A_{f_i}$ & the set of applications allocated to federation $i$\\[0.075cm]
        $a_{f_i}^k$ & the application $k$ allocated to federation $i$\\[0.075cm]
        $usr_j$ & user $j$\\[0.075cm]
        $L(usr_j)$ & the location of user $j$\\[0.075cm]
        $OC(s)$ & the operation costs of server $s$\\[0.075cm]
        $TC(s)$ & the traffic costs of server $s$\\[0.075cm]
        $u(f_i)$ & the utility function of federation $i$\\[0.075cm]
        $Rev(f_i)$ & the revenue of federation $i$\\[0.075cm]
        $\sigma_{a, f_i}$ & the discount factor of application $a$ for federation $i$\\[0.075cm]

        \hline
    \end{tabular}
    \label{table:definitions}
\end{table}

In the previous equation, $\sigma_{a,f_i}$ is the discount factor that adjusts the regular payment made by application $a$ if the federation $f_i$ fails to meet the minimum requirements. Finally, $payment(a)$ represents the regular payment made by application $a$. By summing up the discounted payments of all applications in the federation, we can calculate the revenue expected by that federation.

From the previous, we can deduce that, given a strategy profile $x \in X$, 
that:
\begin{align}
    u(x) = u(F) &= \sum_{f_j \in F} u(f_j) \\
    u_i(x) = u(p_i) &= \sum_{f_j \in F} \frac{ \abs{S_{f_j}^{p_i}} }{ \abs{S_{f_j}} } u(f_j) \nonumber \\
    &= \sum_{f_j \in F} \frac{n_{f_j}^{p_i}}{n_{f_j}} u(f_j) \label{eq:5}
\end{align}
The latter of the previous equations is the utility of the provider $p_i$.

\subsection{Objectives}
\label{section:objectives}
The main objective of this paper is to find a computationally-efficient search algorithm that selects a target strategy profile $x^*$ from the search space $X$ (described in section \ref{section:problem}) for time $t$ such that $x^*$ is efficient and stable. To achieve that, the broker, the entity forming the federations and responsible for performance optimization, should store a model that predicts the performance and load of the network between the fog nodes running. However, that will mean 
that each server has to send its users' data to the broker, which violates the 
privacy of the users. This is where federated learning takes place by securing
the clients' data (described in section \ref{section:methodology_FL}).

After we laid the mathematical model foundation in (section \ref{section:problem}), 
we can now precisely describe the properties required in the target strategy
profile $x^*$. It is required to be efficient and stable. Efficiency is 
formulated as follow:
\begin{align}
    \textit{maximize } u(x^*)
\end{align}
This is equivalent to maximizing utilitarian welfare. Thus, this implies that $x^*$ is Pareto optimal too.
Pareto optimality is a state in which no individual can be made better off without making at least one individual worse off.

Nothing would be relevant unless $x^*$ is stable. Instability would imply 
the existence of an incentive for a fog provider to deviate from its current
strategy, and that would harm the QoS of the federation and thus, its reputation. This could affect the main characteristic of the fog federations, the low latency factor, thus, leading to the dissociation of the whole federation.
\noindent
There are many definitions of stability, the most stringent of them would be 
the strong Nash equilibrium. In this kind of equilibrium, no coalition, fixing the actions of its complements, can profitably deviate from the prescribed strategy profile. However, this kind of equilibrium would be too computationally hard to 
obtain and its existence is not always guaranteed as discussed in \cite{nash}.
\noindent
That limits us to a less strict notion of equilibrium, the strict pure Nash equilibrium, defined as the following:
\begin{equation}
    u_i(x^*_i, x^*_{-i}) > u_i(x_i, x^*_{-i}) \text{ for all } x_i \in X_i \text{ for all } i \label{eq:7}
\end{equation}
The equation states that no player can benefit from a unilateral deviation from their chosen strategy. This choice of stability would come naturally given our choice 
of evolutionary game theory solution as discussed in section \ref{section:methodology_evolutionary}.

We are attempting to find an approximate algorithm to solve this problem because it is an NP-hard problem. This is because it can be reduced to finding a Pareto optimal solution of a Hedonic game with Common Ranking Property. It has been proven that finding such a solution is NP-hard \cite{NPHard}, meaning that an exact algorithm that solves this problem in polynomial time does not exist. Therefore, it is necessary to devise an approximate algorithm that can yield a satisfactory solution with high computational efficiency.

\section{Methodology}\label{section:methodology}
\subsection{Pre-Formation Phase: Prediction of QoS Metrics Through Federated Learning Models}
\label{section:methodology_FL}

\begin{figure}[H]
    \centering
    \includegraphics[width=10cm]{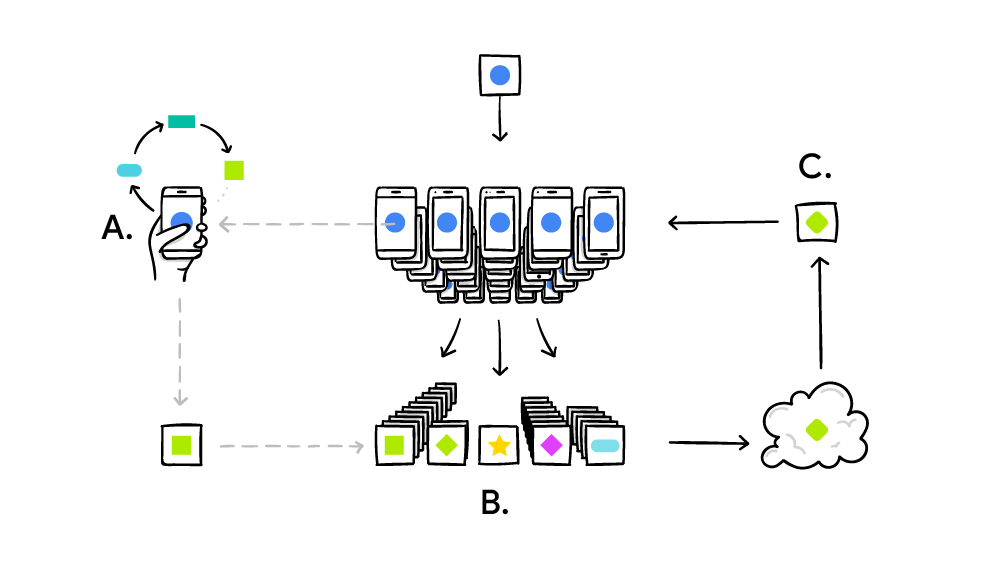}
    \caption{\textmd{Federated Learning Illustration}}
    \label{fig:FL_illustration}
\end{figure}

In order to run the federation formation algorithm, we need to provide it with accurate real-time data about the performance and the load of the network between the requesting user and each of the fog servers. In \cite{Hala}, the authors focused on using a machine-learning model that predicts the QoS values of the requests to the fog nodes. Each fog node sends data to the broker that preprocesses it and then uses it to train the model. However, this model poses a great privacy threat where the broker knows confidential details about the agents' requests. This is why in this paper we are transitioning to a privacy-preserving and decentralized learning scheme, i.e., federated learning.

Federated learning is a decentralized learning scheme that preserves the privacy of the users. Each provider has a local training data set which is never uploaded to the broker.
Instead, each client computes an update to the current global model maintained by the broker, and only this update is communicated. The final model is then obtained on a client by aggregating all the local models into a global model. \cite{FedLearning}
Below is the explanation of the steps required to advance a federated learning model to predict the QoS run-time and throughput for the new requests to the fog servers. We explore the most suitable and practical learning models that yield high levels of accuracy. The following points outline the essential steps required to execute the federated learning architecture:

\textbf{Preprocessing}: Data preprocessing is an essential process that involves cleaning the data and making it suitable for training a machine learning model. It's performed with the goal of increasing the accuracy and efficiency of a machine-learning model. Data preprocessing also involves the process of selecting the features of the input data. Additionally, data preprocessing may include handling missing values, removing outliers, and normalizing or scaling the data to ensure uniformity and improve the performance of the machine learning model. Furthermore, it may involve encoding categorical variables, such as converting text or categorical data into numerical representations, to make them compatible with the algorithms used for training the model

\textbf{Model}: The model lies in the center of the federated learning architecture. Increasing its accuracy is the main goal. The choice of model depends on factors like device capabilities, communication constraints, and the learning task, aiming for a balance between local learning and collaborative model aggregation.
For our problem, we found it suitable to use a regression neural network with multiple convolutional and dense layers.

\textbf{Federated Training}: In Federated Learning architecture, after the global model is shared with each of the nodes, the nodes perform individual training on the device using their local data. Figure \ref{fig:FL_illustration}.A visually depicts this process. This local training process typically involves running an optimization algorithm, such as stochastic gradient descent (SGD), to update the model parameters based on the local data's patterns and features.

\textbf{Federated Averaging}: After the local training, the central server picks a subset of the nodes to share their own local models. Afterward, an averaging algorithm is used to combine the local models into one single global model. This process safeguards the privacy of the nodes by preventing the identification of individual node data within the global model \cite{FL_1}.
This process is called federated averaging as illustrated in Figure \ref{fig:FL_illustration}.B.

\textbf{Evaluation}: Before evaluating or testing the federating learning algorithm, it is imperative to create the testing dataset. For that, there are many methods such as train-test split and k-fold cross-validation. A train-test split divides the dataset randomly into two parts: one for training and another part for testing. Cross-validation is a more robust evaluation model that divides the data into multiple subsets, performing multiple train-test splits. It is more reliable in estimating the model's performance.

Machine learning evaluation metrics are crucial for assessing the performance and effectiveness of a trained model. These methods help in understanding how well a model generalizes to unseen data and its ability to solve the intended task. We now list some of the common testing metrics we used in training our models.
\begin{itemize}
    \item Mean Squared Error (MSE): it measures the average of the square of the differences between the estimated values and the actual value.
    $$ \operatorname {MSE} = {\frac {1}{n}}\sum _{i=1}^{n}\left(Y_{i}-{\hat {Y_{i}}}\right)^{2} $$
    \item Mean Absolute Error (MAE): it measures the absolute value of differences between the predicted and actual values.
    $$ \operatorname {MAE} = {\frac {1}{n}}\sum _{i=1}^{n}\abs{Y_{i}-{\hat {Y_{i}}}} $$
    \item Accuracy: it is a commonly used evaluation metric for categorical machine learning models. It estimates the proportion of correctly classified data points out of the total number in the dataset.

\end{itemize}

The specific implementation details and results are reported in section \ref{section:evaluation_FL}.

\subsection{Federated Learning Client Selection}
\label{section:methodology_client_selection}
Before the process of federated averaging, a subset of federated learning clients is selected for the training cycle. The federated learning client selection process decides which client devices are chosen in each training round. The reasons why this phase is needed is due to the fact that federated learning cannot afford to utilize all of the clients in the same training round. If the number of clients is excessive, it means that a single round would require at least $transfer\_time\_from\_server + train\_time_{f_n} + transfer\_time\_to\_server_{f_n} + agg\_time$ period of time, where $transfer\_time\_from\_server$ is the time needed to transfer the current model from the aggregator to all of the participants of the current round. $train\_time_{f_n}$ is the amount of time required by the slowest provider $f_n$ to finish data training. $transfer\_time\_to\_server_{f_n}$ is the amount of time required by $f_n$ to transfer the model to the aggregator node. $agg\_time$ is the time needed to aggregate all of the received models. Thus, a selection mechanism is required that takes into consideration a certain trade-off of time and learning quality.
Other factors to consider that might hinder the client selection process, including but not limited to, heterogeneous data distributions and system capabilities. In addition, not all clients will be simultaneously available for federated training. Furthermore, including too many clients for the training process may result in inefficient efforts and less-than-optimal performance \cite{fu2023client}.

Hence, a central challenge in practical federated learning lies in the identification of an optimal subset of clients as participants. Each of the clients independently possesses its own private dataset, which gets gathered and aggregated by the federated averaging algorithm.

\textbf{Objectives}: Before delving into the specifics of client selection, it's essential to outline the objectives that guide this process. This section will demonstrate the goals of client selection in federated learning as follows. First and foremost, diversity means ensuring that a diverse set of clients participate to prevent model bias and representativeness issues. This involves considering different types of devices, data distributions, and network conditions. Second, privacy is a main aim that implies protecting the privacy of individual clients by selecting a subset of clients whose data contributions do not risk exposing sensitive information. The selection algorithm also ought to be efficient, selecting clients that can contribute the most useful updates to the global model. Finally, the client selection process needs to be accurate, which means prioritizing clients with high-quality data or models to improve the accuracy of the global model.

\textbf{Strategies}: Various strategies can be employed to select clients for participation in federated learning. The choice of strategy depends on the specific requirements and constraints of the federated learning scenario. Many strategies were enumerated in \cite{fu2023client}. We will mention some basic ones here. One simple strategy is assigning weights to clients based on their data quality, model performance, or other relevant factors. Clients with higher weights are more likely to be selected, improving the convergence and accuracy of the global model. Proximity-based selection is another valid strategy. It works by selecting clients based on their geographical proximity or network latency to the central server. This strategy can help reduce communication overhead and improve training efficiency.

In conclusion, the process of client selection in federated learning is a critical aspect of model training. It involves considering various objectives and strategies to ensure that the global model converges efficiently, accurately, and while preserving privacy.

\subsection{Federation Formation Phase: The Genetic Algorithm}
\label{section:methodology_genetic}
Before discussing stability or federation maintainment, there should exist an initial formation strategy that smooths the execution of the last stage. In this stage, we seek an efficient way to optimally form a set of federations in an unstable environment.

The Genetic algorithm is a metaheuristic commonly used for optimization and search problems. Genetic algorithms were invented by John Holland in the 1960s \cite{GA_Holland_Book}. However, Holland's original goal was to use it to study the phenomenon of adaptation and evolution, when he published his book, Adaptation in Natural and Artificial Systems. Then, Goldberg extended it for optimization and search problems in the 1989s \cite{GA_Goldberg_book}. Since then, Genetic algorithms have been applied in lots of fields such as image and signal Processing, computational finance, bioinformatics, and Robotics. The Genetic algorithm is widely used due to its simplicity and efficiency. They are mainly used to solve resource-demanding problems such as the NP-hard ones. Consideration of the specific problem and appropriate adjustments are essential for the successful application of GAs. The genetic algorithm is founded on three main pillars: (1) the encoding and the search space, (2) the fitness function, and (3) the evolution operators \cite{GA_MIT_book}.

\vspace{0.2cm}
\subsubsection{The Encoding}\hfill\\
In order to apply the genetic algorithm to a given search problem, a specific encoding must be applied. The Genetic algorithm was created to stimulate nature, thus, most of its terminology uses biological terms. We call this encoding a chromosome after the biological carriers of genetic information. Encoding is a crucial step in genetic algorithms as it determines how the information of a solution is structured and represented. Typically, encoding is problem-specific and will defer from one problem to the other. There are multiple encoding techniques that are in use for the Genetic algorithm, including but not limited to binary encoding, integer encoding, permutation encoding, and tree-based encoding. For our problem, we are trying to figure out which federation each fog server should be allocated. Consequently, we used an integer encoding, where a chromosome is a string satisfying the following properties:
\begin{itemize}
    \item The length of the chromosome is the sum of the number of fog servers of all providers:
    $$ \sum_{i=1}^{n} \abs{S_{p_i}}$$
    \item Allowed alphabet $\Gamma$, i.e., the possible symbols allowed in each element of the chromosome is a subset of the integers representing an index of a federation.
    $$ \Gamma = \{ j \in \mathbf{Z} \mid 1 \leq n \leq m \} \subseteq \mathbf{Z}$$
    \item Element $i$ of the chromosome has value $f$ if and only if the fog server indexed $i$ is assigned to federation $f$
\end{itemize}

\noindent
For example, a chromosome \lq1 2 3 1 2 3\rq means that (A) all federations have 6 servers in total, (B) we have 3 fog federations, (C) federation 1 gets the server set $\{ 1, 4 \}$, federation 2 gets the server set $\{ 2, 5 \}$, and federation 3 gets the server set $\{ 3, 6 \}$. Encoding and decoding from and to chromosomes is an essential step for any Genetic algorithm application.

\vspace{0.2cm}
\subsubsection{The Fitness Evaluation}\hfill\\
The Genetic algorithm requires a fitness function in order to discern high-fitness chromosomes from bad ones. The Genetic algorithm runs the fitness function on every chromosome to give it a fitness score. In our problem, we can use the utility function of the federations in order to model the fitness function $\mathfrak{F}$. Thus, given a chromosome $c$, we decode it to get the sets of federations $f_i \leftarrow \operatorname{decode}(c)$. After that, it is evident that the fitness function should be dependent on the utilities of the federations.
$$\mathfrak{F}(c) = G\big( u(f_1), u(f_2), \dots, u(f_m) \big)$$
It is only left to find the function $G$ experimentally, however, it's intuitive that it needs to incorporate both (1) the summation of the utility functions to achieve a high payoff solution, and (2) the variance of the utility function in order to achieve fairness between different providers.

\vspace{0.2cm}
\subsubsection{The Evolution Operators}\hfill\\
The genetic algorithm contains three types of operators: selection, crossover, and mutation. The first operator is selection. Before proceeding to the next generation, the Genetic algorithm selects which chromosomes will be carried over. This process micks evolutionary pressure as by selecting individuals with better fitness, the genetic algorithm promotes the propagation of traits and characteristics that contribute to improved solutions over time. There are many methods of selection. Elitist selection only selects the top-ranked chromosomes. Roulette Wheel Selection, which is used in our implementation, selects a chromosome with a probability based on its relative fitness. Roulette Wheel Selection is better than the former method as it encourages exploration and diversity \cite{GA_selection}.

The crossover operation begins after the selection operation is performed. It involves combining genetic material from parent individuals to create offspring with potentially improved characteristics. The crossover operation involves randomly picking a crossover point in the parent chromosomes, and then exchanging the genetic material of the parents at this point to produce two offspring. For example, to crossover \lq 1 2 3 4 5\rq with \lq a b c d e\rq, a random crossover point is first picked, let's say index 2. Then, the exchanging process happens to produce \lq 1 2 c d e\rq with \lq a b 3 4 5\rq.
The crossover operation in genetic algorithms facilitates the exploration of the solution space by combining genetic information from parent individuals to create diverse offspring \cite{GA_crossover}.

The mutation operation serves as another mechanism to introduce diversity into the population and is the most straightforward of the three genetic operators. By randomly selecting an element of the chromosome, it alters that element with a certain probability. While crossover primarily focuses on combining genetic material from parent individuals, mutation introduces small, random changes to individual chromosomes, allowing for exploration beyond the existing solutions

\vspace{0.2cm}
\subsubsection{GA Implementation}\hfill\\
Algorithm \ref{alg:genetic} illustrates the pseudocode of the genetic algorithm. Lines 1-4, represent the initialization process. The algorithm evaluates every chromosome and then sorts them by fitness. Lines 5-18, represent a loop that represents the evolution of the population. It only terminates when the end condition is met. In the first step of the loop (lines 6-10), a bunch of genetic operators are being applied to the population. First, crossover operation between every two chromosomes, followed by a mutation. The resultant chromosomes are added to the gene pool. Upon termination, the algorithm will return the first element which is the fittest.

\begin{algorithm}
    \caption{Genetic Algorithm Pseudocode}
    \label{alg:genetic}

    $t \gets 0$ \\
    $\operatorname{initialize\_population}(P)$ \\
    $\operatorname{evaluate}(P)$ \\
    $\operatorname{sort\_by\_fitness}(P)$ \\
    \Loop
    {
        \For {$i, j \in P$}
        {
            $i', j' \gets \operatorname{crossover}(i, j)$ \\
            $\operatorname{mutate}(i')$ \\ 
            $\operatorname{mutate}(j')$ \\
            $P \gets P \cup \{i', j'\}$
        }
        $\operatorname{evaluate}(P)$ \\
        $\operatorname{sort\_by\_fitness}(P)$ \\
        $\operatorname{select}(P)$ \\
        \If {end condition is met}
        {
            break
        }
        $t \gets t+1$ \\
    }
    \Return $P[0]$
\end{algorithm}

\subsection{Federation Formation: Evolutionary Game Theory}
\label{section:methodology_evolutionary}

\begin{figure}[t]
\center
    \begin{tikzpicture}[node distance=1.5cm, every node/.style={font=\sffamily}, align=center, every node/.style={scale=0.8}]
      \node (initial)           [process]    {Initial Population};
      \node (eval)              [process, below of=initial]           {Evaluate Strategies};
      \node (decision)          [decision, aspect=2, below of=eval, yshift=-0.5cm]           {Satisfactory \\ Solution?};
      \node (yes)               [process, right of=decision, xshift=2.5cm]           {\text{  End  }};
      \node (nono)              [process, left of=decision, xshift=-2.5cm]             {Population n+1};
      \node (no)                [process, below of=decision, yshift=-1cm]           {Replicate Successful \\Strategies};
      \draw[->]             (initial) -- (eval);
      \draw[->]             (eval) -- (decision);
      \draw[->]             (decision) -- node[midway,fill=white] {No} (no);
      \draw[->]             (no) -- (no -| nono) -- (nono);
      \draw[->]             (nono) -- (eval -| nono) -- (eval);
      \draw[->]             (decision) -- node[midway,fill=white] {Yes} (yes);
    \end{tikzpicture}
    \caption{\textmd{Evolutionary game theory flowchart}}
    \label{fig:flowchart}
\end{figure}

\subsubsection{Background}\hfill\\
Forming stable fog federations efficiently is 
the main goal of this work. By stability, we mean that the underlying infrastructure is not changing due to providers' uncertainty about their selected federation to join.
As indicated previously, we are interested in searching for the optimal strategy profile $x^* \in X$.

Game theory is the study of mathematical models involving strategic rational agents. Game theory provides a framework to understand and predict behaviors in various fields, such as economics, politics, biology, and computer science, and has practical applications in areas like negotiation, pricing, and resource allocation.
Evolutionary game theory is the extension of game theory using the Darwinian theory of evolution. In evolutionary game theoretic models, agents do not have to be rational about their decisions. Instead, one of the main features of evolutionary game theory is the focus on the dynamics of the strategy changes, and which strategy can dominate other strategies. Individuals with more fit strategies will be replicated more frequently than others. 
This approach allows for the exploration of how different strategies can emerge and evolve over time through processes such as selection, mutation, and replication. By studying the relation between strategy dynamics and evolutionary forces, evolutionary game theory provides insights into the long-term outcomes and stability of strategic interactions in complex systems.

The goal behind evolutionary game theory is to reach the evolutionary stable strategy (ESS). The ESS is the strategy that if almost all agents adopt it, then the fitness
of those members should be higher than any invading mutation. A strategy $x$ is ESS: if under an invasion of strategy $y$, the invaders are forced to adopt the strategy $x$ after a few generations \cite{ESS}.
For this reason, if a strategy $x^*$ is ESS, it's a strict Nash equilibrium as given in Equation \eqref{eq:7}.

Several game theoretic models including hedonic, coalitional/cooperative, non-cooperative, and evolutionary models can be modeled to address the stability of the population. The main reason behind choosing evolutionary game theoretic models was that in simpler models found in literature, evolutionary game theoretic models outperform all the aforementioned algorithms in stability and QoS \cite{GTMainFog}. Furthermore, genetic algorithms were excluded from consideration since they are incapable of forming stable coalitions, which could adversely impact profits and reputation, in the long run \cite{GTMainCloud}.

\subsubsection{Game Elements}\hfill\\
In this subsection, we present the elements of an evolutionary game theoretic model, how to reach an evolutionary stable strategy (ESS), and how to determine whether an ESS has been reached within a population.

An evolutionary game consists of three main parts: (1) the players, (2) strategies, and (3) utility function.
The components of our evolutionary game model are the following:
\begin{itemize}
    \item \textbf{Players}: in our architecture, those are the fog providers. Clearly, they are the ones that make the decision of joining a federation.
the decision-makers.
    \item \textbf{Strategy}: generally speaking, a strategy is the actions portfolio a player chooses to abide by at every decision point. In our architecture, every player $p_i$ can select a strategy $x_i$ from the strategy space described before $X_i$. This corresponds to a fog provider picking a partition of servers to allocate to some federations.
    \item \textbf{Utility Function}: the utility function determines the fitness of each player. Thus, it determines how likely is it to be replicated in the next round. In our architecture, each player $p_i$ has the utility function given in equation \eqref{eq:5} by $u_i(x)$
\end{itemize}

Figure \ref{fig:flowchart}, represents a flowchart of the evolutionary algorithm used to obtain an ESS. After evaluating the fitness of each player, we check if an ESS is reached. If not, we replicated successful strategies and continue till termination. The population fitness will keep improving its strategies till equilibrium, equivalently the optimal strategy $x^*$ is achieved.

Next, we discuss the replicator equation, a differential equation that quantifies the evolution of strategies over time. The folk theorem of evolutionary game theory makes the replicator equation really powerful, as it formalizes a relation between the solutions of the replicator equation, and nash equilibria \cite{replicator}. The replicator equation describes the dynamics of strategy frequencies in a population over time. It quantifies how the proportions of different strategies change based on their relative payoffs or fitness values.

Let us consider now a population consisting of $n$ strategies, and let $x_i$ be the frequency of strategy $i$, $x$ be a vector of distribution of the strategies. Let $f_i(x)$ be the fitness of the $i$th player, and $v(x)$ be the population average fitness. We can conclude that
$$ \sum_{i=1}^{n} x_i = 1, $$
$$ v(x) = \sum_{i=1}^{n} x_i f_i(x) $$
The replicator equation for player $i$ is presented as follows:
$$ \dot x_i = x_i \cdot (f_i(x) - v(x)) $$

Translating this to our specific problem, we can substitute $f_i$ with our utility function $u_i$, and evaluate whether a strategy distribution $x$ satisfies the replicator function.
This condition serves as the termination condition equivalent to identifying an ESS in our game.

\section{Experimental Evalutation}
\label{section:evaluation}

\subsection{Setup of the Federated Learning Model}
In this section, we present the specifics of the federated learning environment step. 

\textbf{Data}: we used WS-Dream data set\footnote{https://wsdream.github.io/} which is a Distributed REliability Assessment Mechanism for Web services. It contains a web services QoS prediction component. This data set contains a real-world QoS evaluation results from 339 users on 5825 web services. It provides evaluation for both run-time and throughput. We used these to train two federated learning models one for each metric. 

The features we were interested in were User-Id, User-Longitude, User-Latitude, Fog-Node-Id, Fog-Node-Longitude, Fog-Node-Latitude, Invocation-Throughput, and Invocation-Response-Time. The carefully chosen features establish a strong correlation among various factors, contributing to the achievement of our desired predicted throughput and response time.

\textbf{Model}: for the learning model, we choose a neural network model with multiple convolutional and dense layers. We implemented the model using TensorFlow and exported it to TensorFlow federated, which is an open-source framework for federated learning. It allows for privacy-preserving, collaborative learning across multiple devices or servers.

\textbf{Preprocessing}: We employed one-hot encoding to encode the User-Id and Fog-Node-Id while applying min-max normalization to the throughput and the run-time. Furthermore, we mapped the coordinate values of latitude and longitude to the unit interval.

\subsection{Evaluation of the Federated Learning Models}
\label{section:evaluation_FL}

\begin{figure}[h]
    \centering
    \includegraphics[width=1\linewidth]{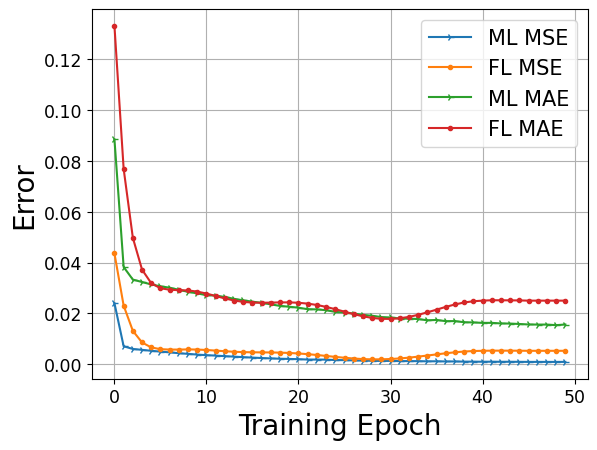}
    \caption{\textmd{Training Metrics of the Response Time Model}}
    \label{fig:model_rt}
\end{figure}
\begin{figure}[h]
    \centering
    \includegraphics[width=1\linewidth]{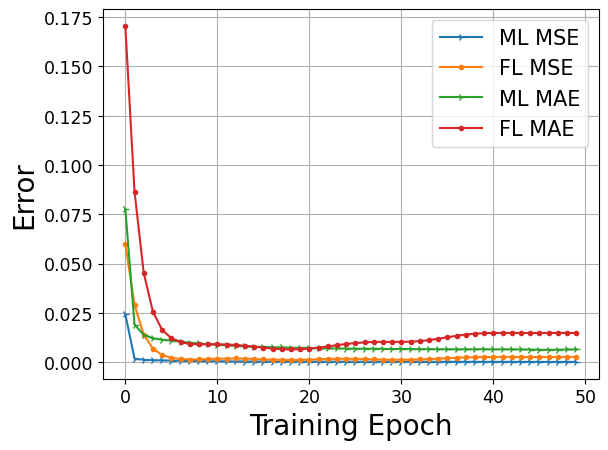}
    \caption{\textmd{Training Metrics of the Throughput Model}}
    \label{fig:model_tp}
\end{figure}

\label{section:evaluation_FL}
In this subsection, we will evaluate our federated learning model against a baseline machine learning model trained on the same data set. In addition, we will compare the federated learning model's final metrics to the machine learning model trained on the same data set used in Shamseddine et al \cite{Hala}.

For evaluating the QoS prediction model, we use two evaluation metrics to judge the federated learning model: mean squared error (MSE), and mean absolute error (MAE). These metrics were described in detail in section \ref{section:methodology_FL}.

The federated learning model can't perform significantly better than the machine learning model due to the nature of the training. Unlike the ML models where the training could happen on the whole data set, the FL models' training happens in updates involving a small subset of the nodes. Figures \ref{fig:model_rt} and \ref{fig:model_tp} show the metrics of the federated learning model evolving during the training. This is plotted against a machine learning model trained on the same data. We can see clearly that the error metrics of the FL don't vary significantly when contrasted with the ML models. That result shows that replacing the ML with the FL models would have almost the same accuracy and will not affect our fog federation formation protocol negatively.

\begin{table}[H]
    \centering
    \caption{\textmd{Testing Metrics for the FL and ML Models}}
    \begin{tabular}{| c ||c | c | c | c |} 
        \hline
        & \multicolumn{2}{|c|}{Response Time} &  \multicolumn{2}{|c|}{Throughput} \\
        \hline
        & MSE & MAE & MSE & MAE \\
        \hline\hline
        FL & 0.00521 & 0.02497 & 0.00266 & 0.01487 \\ 
        ML & 0.00084 & 0.01557 & 0.00018 & 0.00649 \\
        \hline
    \end{tabular}
    \label{table:testing_metrics}
\end{table}

For evaluating the testing metrics, we check table \ref{table:testing_metrics}. We notice that for the response time the MSE and MAE are almost the same for both the FL and ML models. 
In terms of throughput, it can be observed that the ML model has lower error rates compared to the FL model, although the difference is not significant.

We also compare the testing results to the results of the Bagging model in \cite{Hala} which yielded the best results in that paper. The bagging method applied to the response time data set obtained an MAE of $0.0203$ which is comparable to our FL model ($MAE = 0.02497$). 
Our FL model demonstrated better performance in terms of throughput analysis, with an MAE of $0.01487$, outperforming the bagging method described in Shamseddine et al \cite{Hala}, which had an MAE of $0.659$.

\subsection{Evaluation of the Client Selection Process}
In this section, we assess the federated learning model equipped with a client selection algorithm as described in section~\ref{section:methodology_client_selection}. As a baseline, we use the federated learning model evaluated in the previous subsection. For evaluating the two QoS prediction models, we use two evaluation metrics, the mean squared error (MSE), and the mean absolute error (MAE).

We noticed using the WSDream dataset that the clients have a significant variation in the size of their individual datasets. This observation prompted the adoption of a weighted selection algorithm, wherein the likelihood of selection becomes contingent on the size of the client's dataset.

We noticed an increase in the federated learning model's accuracy once we implemented the previously mentioned client selection step. For the throughput models, we noticed an average $70.63\%$ reduction of the MSE metric, and $36.46\%$ reduction of the MAE metric.

\begin{figure}[H]
    \centering
    \begin{subfigure}[t]{0.49\textwidth}
        \centering
        \includegraphics[width=\linewidth]{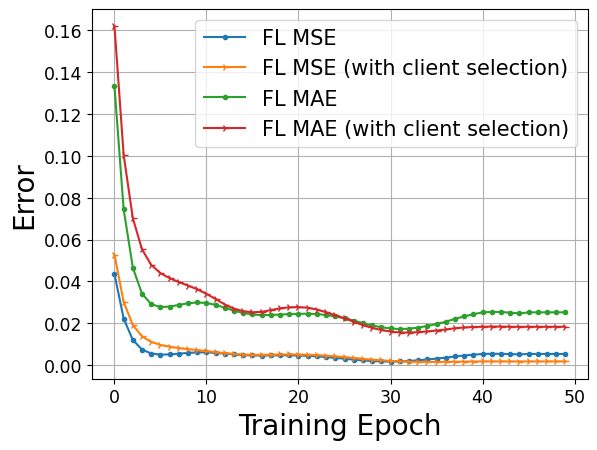}
        \caption{\textmd{Response Time Models}}
        \label{fig:client_rt}
    \end{subfigure}
    \begin{subfigure}[t]{0.49\textwidth}
        \centering
        \includegraphics[width=\linewidth]{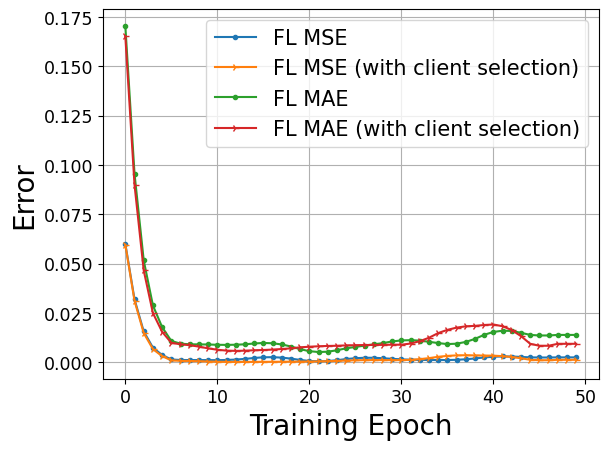}
        \caption{\textmd{Throughput Models}}
        \label{fig:client_tp}
    \end{subfigure}
    \caption{\textmd{Federated Learning Model Metrics Comparison with and without Client Selection}}
\end{figure}

Figures~\ref{fig:client_rt}~and~\ref{fig:client_tp} show the metrics of the federated learning model evolving during the training. This is plotted against the same model but with the inclusion of client selection.
We can see on average after enough training rounds, the error metrics of the model are lower when we use the client selection step. Intuitively, we rely on the model with client selection to be our QoS assessor for the next two stages.

\subsection{Evaluation of the Fog Federation Formation}
\label{section:evaluation_formation}

\begin{figure*}[!t]
     \centering
     \begin{subfigure}[t]{0.49\textwidth}
        \centering
        \includegraphics[width=\linewidth]{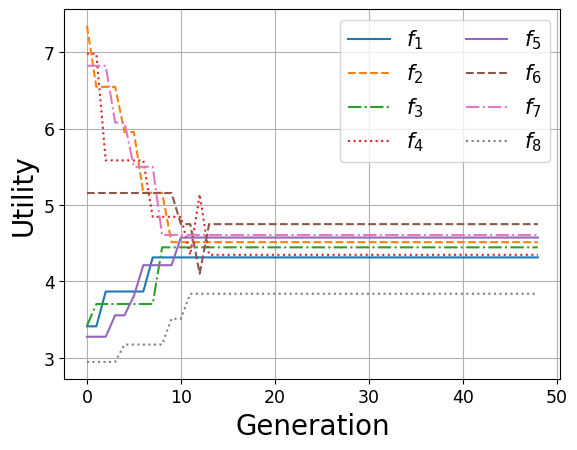}
        \caption{\textmd{Evolutionary Game Theory Approach}}
        \label{fig:form_evol}
     \end{subfigure}
     \begin{subfigure}[t]{0.49\textwidth}
        \centering
        \includegraphics[width=\linewidth]{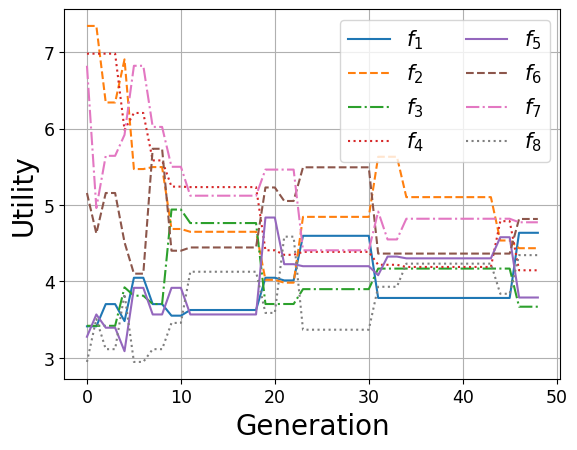}
        \caption{\textmd{Genetic}}
        \label{fig:form_genetic}
     \end{subfigure}
     \hfill
     \begin{subfigure}[t]{0.49\textwidth}
        \centering
        \includegraphics[width=\linewidth]{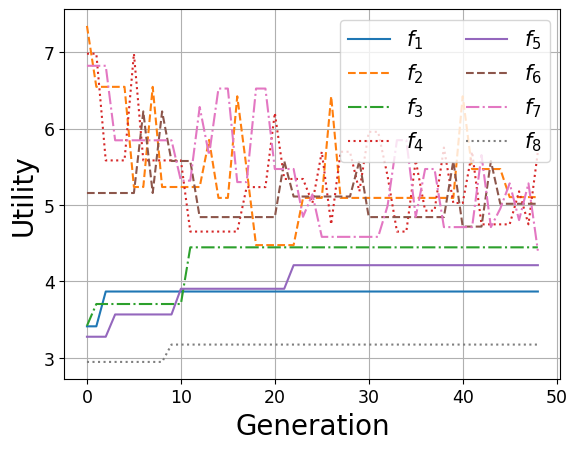}
        \caption{\textmd{Greedy Approach}}
        \label{fig:form_greedy}
     \end{subfigure}
     \caption{\textmd{Comparison of the Stability of Different Formation Techniques}}
\end{figure*}

In this section, we will evaluate our adaptive federation formation model in terms of (1) stability, (2) quality of service, and (3) federations payoff. We compare our approach to two baselines: (1) the genetic approach presented in Shamseddine et al \cite{Hala}, 
and (2) a greedy approach.
The greedy algorithm is almost identical to our algorithm except that it's executed simultaneously by each player without coordination. At every round, every player moves to the federation that provides the local maximum, hence it's greedy. We will be using the same federated learning model as a quality of service predictor for the three approaches. For the Genetic and the Greedy approaches, we assign the initial population using K-means and let each approach update the population using its own method.

\subsubsection{Stability}\hfill\\
Figures \ref{fig:form_evol}-\ref{fig:form_greedy} show how an initial population develops over generations using the three different approaches we considered. The x-axis represents the timeline. The y-axis represents the utility of each of the 8 federations in this example. We can notice when $x<15$, the evolutionary approach is converging quickly to an ESS. When $x=15$, stability is achieved and persists till the end. This is attributed to the convergence properties of the evolutionary algorithm as discussed in section \ref{section:methodology_evolutionary}. On the other hand, the other two approaches could not stabilize at all. 
Such occurrences lead providers to switch between federations, resulting in utility losses and dissatisfaction among users.

\subsubsection{Quality of Service}\hfill

\begin{figure}[H]
    \centering
    \includegraphics[width=1\linewidth]{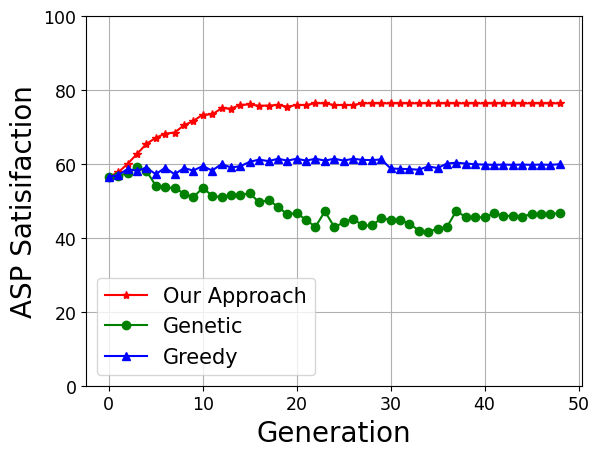}
    \caption{\textmd{User Satisfaction Percentage of the Response Time}}
    \label{fig:satisfaction_rt}
\end{figure}
\begin{figure}[H]
    \centering
    \includegraphics[width=1\linewidth]{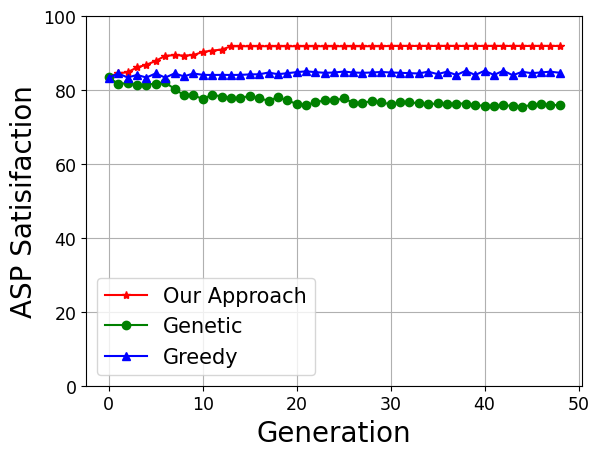}
    \caption{\textmd{User Satisfaction Percentage of the Throughput}}
    \label{fig:satisfaction_tp}
\end{figure}


Figure \ref{fig:satisfaction_rt} describes the percentage of users satisfied by the response time in the three approaches. We observe that our scheme was capable of reaching up to 78\%, while the other approaches fell behind. Figure \ref{fig:satisfaction_tp} shows the percentage of users satisfied by the throughput. Again, the evolutionary approach tops the benchmark methods, approaching 92\%. In both scenarios, it is observed that the graph for the alternative approaches lacks monotonicity, indicating their lack of reliability.

\subsubsection{Federations Payoff}\hfill\\
We also evaluate the three approaches based on the total federations' payoff. The simulation results showed that our approach, the one empowered by evolutionary game model, outperformed the other two approaches, achieving a total federation payoff of \$380, compared to \$360 and \$355 for the genetic and greedy approaches respectively (rounded to the nearest \$5). This indicates that our approach not only offers greater stability but also yields the highest federations' payoff.




\section{Conclusion}
\label{section:conclusion}

Fog federation has gained recent interest since it helps improve the QoS of fog providers, and thus eliminates the bottleneck in developing IoT applications.
However, fog federations may suffer from instabilities that might incentivize some providers to leave the federation. That would harm the QoS of the federation. In this paper, we proposed a decentralized architecture for forming stable fog federations.
It consists of a dynamic QoS prediction model that has twofold benefits over the previous model. On one hand, it eliminates the costs associated with the broker. On the other hand, it eliminates data privacy threats by utilizing federated learning models.
Utilizing the prediction model, an initial fog federation formation scheme is devised using a Genetic Algorithm, followed by an evolutionary game theoretical model to stabilize the federations. 
The numerical findings demonstrate that the formation process reaches a state of stability, leading to enhanced payoff and quality of service in terms of response time and throughput.

\section*{Acknowledgment}
This work was supported in part by the Natural Sciences and Engineering Research Council of Canada (NSERC).

\printbibliography

\noindent \textbf{Biography}



\begin{IEEEbiographynophoto}{Zyad Yasser}
graduated from the New York University Abu Dhabi in 2023 with a Bachelor of Science degree in Computer Science. His research interests include Cloud Computing, Fog Computing, Cloud and Fog Federations, Resource Management, Artificial Intelligence, and Game theory.\
\end{IEEEbiographynophoto}

\begin{IEEEbiographynophoto}{Ahmad Hammoud}
received the B.S. degree in business computing from Lebanese University in 2016, the M.S. degree in computer science from Lebanese American University in 2019, and the Ph.D. degree from École de Technologie Supérieure (ETS), Canada in 2023. He is a PostDoc Fellow at ETS. His current research interests include Metaverse, cloud and fog federations, federated learning, Internet of Vehicles, game theory, blockchain, artificial intelligence, and security
\end{IEEEbiographynophoto}

\begin{IEEEbiographynophoto}{Azzam Mourad}
received his M.Sc. in CS from Laval University, Canada (2003) and Ph.D. in ECE from Concordia University, Canada (2008). He is currently a Visiting Professor at Khalifa University, a Professor of Computer Science and Founding Director of the Artificial Intelligence and Cyber Systems Research Center at the Lebanese American University, and an Affiliate Professor at the Software Engineering and IT Department, École de Technologie Supérieure (ETS), Montreal, Canada. His research interests include Cyber Security, Federated Machine Learning, Network and Service Optimization and Management targeting IoT and IoV, Cloud/Fog/Edge Computing, and Vehicular and Mobile Networks. He has served/serves as an associate editor for IEEE Transactions on Services Computing, IEEE Transactions on Network and Service Management, IEEE Network, IEEE Open Journal of the Communications Society, IET Quantum Communication, and IEEE Communications Letters, the General Chair of IWCMC2020-2022, the General Co-Chair of WiMob2016, and the Track Chair, a TPC member, and a reviewer for several prestigious journals and conferences. He is an IEEE senior member.
\end{IEEEbiographynophoto}

\begin{IEEEbiographynophoto}{Hadi Otrok}
received his Ph.D. in ECE from Concordia University. He holds a Full Professor position in the Department of Computer Science (CS) at Khalifa University. Also, he is an Affiliate Associate Professor at the Concordia Institute for Information Systems Engineering at Concordia University, Montreal, Canada, and an Affiliate Associate Professor in the Electrical Department at Ecole de Technologie Superieure (ETS), Montreal, Canada. His research interests include the domain of blockchain, reinforcement learning, crowd sensing and sourcing, ad hoc networks, and cloud security. He co-chaired several committees at various IEEE conferences. He is also an Associate Editor at IEEE Transactions on Services Computing, IEEE Transactions on Network and Service Management (TNSM), Ad-hoc Networks (Elsevier), and IEEE Network. He also served from 2015 to 2019 as an Associate Editor at IEEE Communications Letters.
\end{IEEEbiographynophoto}

\begin{IEEEbiographynophoto}{Zbigniew Dziong}
 received the Ph.D. degree from the Warsaw University of Technology, Poland, where he also worked as an Assistant Professor. From 1987 to 1997, he was with INRS-Telecommunications, Montreal, QC, Canada. From 1997 to 2003, he worked with Bell Labs, Holmdel, NJ, USA. Since 2003, he has been with the École de Technologie Supérieure (University of Quebec), Montreal, as a Full Professor. He is an expert in the domain of performance, control, protocol, architecture, and resource management for data, wireless, and optical networks. He has participated in research projects for many leading telecommunication companies, including Bell Labs, Nortel, Ericsson, and France Telecom. He won the prestigious STENTOR Research Award (1993, Canada) for collaborative research. His monograph ATM Network Resource Management (McGraw Hill, 1997) has been used in several universities for graduate courses.
\end{IEEEbiographynophoto}

\begin{IEEEbiographynophoto}{Mohsen Guizani}
received the BS (with distinction), MS and PhD degrees in Electrical and Computer engineering from Syracuse University, Syracuse, NY, USA in 1985, 1987 and 1990, respectively. He is currently a Professor of Machine Learning and the Associate Provost at Mohamed Bin Zayed University of Artificial Intelligence (MBZUAI), Abu Dhabi, UAE. Previously, he worked in different institutions in the USA. His research interests include applied machine learning and artificial intelligence, Internet of Things (IoT), intelligent systems, smart city, and cybersecurity. He was elevated to IEEE Fellow in 2009 and was listed as a Clarivate Analytics Highly Cited Researcher in Computer Science in 2019, 2020 and 2021. Dr. Guizani has won several research awards including the ‘‘2015 IEEE Communications Society Best Survey Paper Award’’, the Best ComSoc Journal Paper Award in 2021 as well five Best Paper Awards from ICC and Globecom Conferences. He is the author of ten books and more than 800 publications. He is also the recipient of the 2017 IEEE Communications Society Wireless Technical Committee (WTC) Recognition Award, the 2018 AdHoc Technical Committee Recognition Award, and the 2019 IEEE Communications and Information Security Technical Recognition (CISTC) Award. He served as the Editor in-Chief of IEEE Network and is currently serving on the Editorial Boards of many IEEE Transactions and Magazines. He was the Chair of the IEEE Communications Society Wireless Technical Committee and the Chair of the TAOS Technical Committee. He served as the IEEE Computer Society Distinguished Speaker and is currently the IEEE ComSoc Distinguished Lecturer.
\end{IEEEbiographynophoto}

\end{document}